\documentclass[lettersize,journal]{IEEEtran}
\usepackage{amsmath,amsfonts}
\usepackage{algorithmic}
\usepackage{algorithm}
\usepackage{array}
\usepackage[caption=false,font=normalsize,labelfont=sf,textfont=sf]{subfig}
\usepackage{textcomp}
\usepackage{stfloats}
\usepackage{url}
\usepackage{verbatim}
\usepackage{graphicx}
\usepackage{cite}

\usepackage{multirow}

\begin{document}

\title{ISTA-Inspired Network for Image Super-Resolution}

\author{Yuqing Liu$^{1}$, Wei Zhang$^{1,*}$, Weifeng Sun$^2$, Zhikai Yu$^1$, Jianfeng Wei$^1$ and Shengquan Li$^1$\\
$^1$ Pengcheng Laboratory, Shenzhen, China.\\
$^2$ Dalian University of Technology, Dalian, China.


}



\maketitle

\begin{abstract}
    Deep learning for image super-resolution (SR) has been investigated by numerous researchers in recent years. Most of the works concentrate on effective block designs and improve the network representation but lack interpretation. There are also iterative optimization-inspired networks for image SR, which take the solution step as a whole without giving an explicit optimization step. This paper proposes an unfolding iterative shrinkage thresholding algorithm (ISTA) inspired network for interpretable image SR. Specifically, we analyze the problem of image SR and propose a solution based on the ISTA method. Inspired by the mathematical analysis, the ISTA block is developed to conduct the optimization in an end-to-end manner. To make the exploration more effective, a multi-scale exploitation block and multi-scale attention mechanism are devised to build the ISTA block. Experimental results show the proposed ISTA-inspired restoration network (ISTAR) achieves competitive or better performances than other optimization-inspired works with fewer parameters and lower computation complexity.
\end{abstract}

\section{Introduction}
Image super-resolution (SR), as one of the traditional image restoration tasks, has been widely investigated by researchers~\cite{cin1, cin2}. Given a low-resolution (LR) image, the task of image SR is to restore a corresponding high-resolution (HR) instance with more details. There are numerous applications considering the image SR, such as video deinterlacing~\cite{liu2021_icme} and compression~\cite{compression}, remote sensing~\cite{tuo2021_igrass, cin_remote, cin_radar}, EGG analysis~\cite{cin_egg}, and spatiospectral analysis~\cite{ma2022_tci}.

Deep learning has demonstrated its amazing performance in image restoration. There are numerous convolutional neural networks (CNNs) specially designed for image SR. SRCNN~\cite{srcnn} is the first CNN-based method for image SR. After that, deeper and wider networks show their effectiveness with better performance, such as VDSR~\cite{vdsr}, EDSR~\cite{edsr}, RDN~\cite{rdn} and RCAN~\cite{rcan}.  Recent image SR networks usually develop effective blocks for improving the network representation. IMDN~\cite{imdn} and EFDN~\cite{rfdn} utilize information distillation mechanisms to build an efficient network for fast and accurate image SR. Cross-SRN~\cite{crosssrn} builds an edge-preserving network with cross convolution. However, these works concentrate on the block designs but lack interpretation, which limits the performance.

Since image SR can be regarded as a classical optimization task~\cite{ircnn}, there are also works considering building the image SR network from the optimization perspective. IRCNN~\cite{ircnn} provides an iterative solution for the general image restoration task and designs a CNN-based network to solve the denoising prior. ISRN~\cite{isrn} develops an iterative network with the help of the half-quadratic splitting (HQS) strategy. DPSR~\cite{dpsr} and USRNet~\cite{usrnet} also achieve good performance on image SR inspired by the HQS strategy. There are also works building the network by alternating direction method of multipliers (ADMM). Plug-and-Play ADMM~\cite{pap_admm} regards the denoiser as a network prior for different image restoration tasks. ADMMNet~\cite{admmnet} provides an end-to-end network for the compression sensing task. PSRI-Net~\cite{admm2} considers ADMM for SAR image SR. Although these works provide an interpretable network design, the CNN architectures just task the solution step as a whole, without giving an explicit optimization step on how to solve the denoising problem.

\begin{figure}
    \centering
    \includegraphics[width=\linewidth]{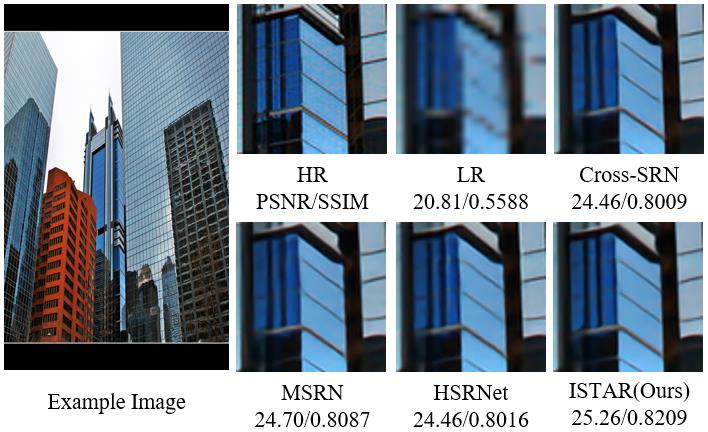}
    \caption{An example comparison among different image super-resolution methods.}
    \label{fig:slogan}
\end{figure}

In this paper, we develop an unfolding network based on the iterative shrinkage thresholding algorithm (ISTA). Different from designing the CNN to directly solve the optimization step, ISTA blocks are specially designed to conduct the image restoration following the ISTA steps. In the ISTA block, CNNs are utilized to adaptively learn the functions in the feature space and speed up the optimization steps. To improve the network representation, multi-scale exploration (MSE) and multi-scale attention (MSA) mechanisms are utilized to build the ISTA block. An ISTA-inspired restoration network (ISTAR) is developed based on the ISTA block for effective image SR. Experimental results show the proposed ISTAR can achieve competitive or better performance than other works.  Compared with other optimization-inspired methods, ISTAR achieves better performances with much fewer parameters and lower computation complexity. Figure~\ref{fig:slogan} shows an example comparison among different image super-resolution methods. Compared with state-of-the-art methods, our proposed ISTAR can generate more satisfying textures that close to the HR image.

The contributions of this paper can be concluded as follows:
\begin{itemize}
	\item We analyze the image super-resolution task from the optimization perspective and develop an ISTA block for image super-resolution.
	\item We develop the multi-scale exploration and multi-scale attention mechanism in the ISTA block, which improves the network representation and boosts the performance.
	\item Experimental results show the proposed network achieves competitive or better performance than other optimization-based works with much fewer parameters and lower computation complexity.
\end{itemize}

\section{Related Works}
\subsection{Deep Learning for Image Super-Resolution}
Deep learning has demonstrated its amazing performance on various computer vision tasks. There are numerous convolutional neural networks (CNNs) specially designed for image super-resolution (SR). SRCNN~\cite{srcnn} is the first CNN-based image SR method composed of three convolution layers, which follows a sparse-coding manner. After SRCNN, deeper and wider networks has proposed to improve the restoration performance. FSRCNN~\cite{fsrcnn} increases the network depth and decreases the input resolution, which makes the method faster and more effective. VDSR~\cite{vdsr} develops a very deep network with residual connection to restore the high-resolution (HR) images. EDSR~\cite{edsr} then utilizes the residual blocks in the network and improves the network capacity. ESPCN~\cite{espcn} provides a different upsampling strategy to restore the HR images, which is more effective than the deconvolution operation. Recently, researchers concentrate more on effective block design for better restoration performance. RDN~\cite{rdn} combines the residual connection~\cite{resnet} and densely connection~\cite{densenet}, and develops a residual dense block for image SR. After that, the researchers introduce the residual-in-residual design with channel attention~\cite{senet} for image SR and build an effective network termed RCAN~\cite{rcan}. RFANet~\cite{rfanet} expands the residual connection and aggregates the residual features for better information transmission. IMDN~\cite{imdn} and RFDN~\cite{rfdn} build the lightweight networks with the help of an information distillation mechanism. SHSR~\cite{shsr} and MSRN~\cite{msrn} utilize hierarchical exploration to further investigate the image features. These works usually concentrate on the effective block designs but neglect to analyze the image SR from the optimization perspective.

\subsection{Optimization-Inspired Image Super-Resolution}
There are also optimization-inspired networks for interpretable image SR. ADMM-Net~\cite{admmnet} provides a good example of dealing with the image restoration problem by the optimization strategy and develops a CNN-based denoiser for plug-and-play restoration. IRCNN~\cite{ircnn} then analyzes the image restoration with the help of the half-quadratic splitting (HQS) strategy and recovers the image with a CNN-based denoiser prior. After IRCNN, there are numerous HQS-based methods for effective image SR. DPSR~\cite{dpsr} proposes a different observation model for image SR and uses kernel estimation and CNN denoiser for plug-and-play image SR. USRNet~\cite{usrnet} develops an end-to-end network for different image SR tasks. ISRN~\cite{isrn} devises an effective network for image SR under the guidance of HQS and maximum likelihood estimation (MLE). HSRNet~\cite{hsrnet} also investigates the HQS strategy and develops a network for aliasing suppression image SR. However, these works just take the solution as a whole and calculate it directly by CNNs, without giving an explicit optimization step for each iteration.

\begin{figure*}[t]
	\centering
	\includegraphics[width=0.8\linewidth]{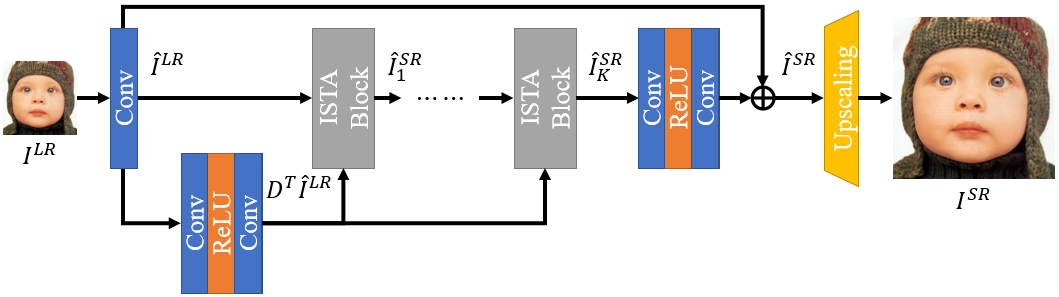}
	\caption{Network design of ISTAR}
	\label{fig:network}
\end{figure*}

\section{Methodology}
In this section, we first analyze the image super-resolution (SR) from the optimization perspective and propose an iterative solution with the help of ISTA. Then, we introduce the designed end-to-end network ISTAR. After that, we discuss the design of the ISTA block. Finally, the network settings are described in detail.

\subsection{ISTA for Image Super-Resolution}
Given an low-resolution (LR) image $I^{LR}$, the task of image SR is to find a corresponding image $I^{SR}$, satisfying
\begin{equation}
	\label{eq1}
	I^{SR} = \arg\min_{I^{LR}}||DI^{SR}-I^{LR}||_\ell^2+\lambda ||I^{LR}||_1,
\end{equation}
where $D$ is the down-sampling matrix, and $\lambda$ is a weighting factor. The prior term $\lambda||I^{LR}||_1$ is utilized to introduce the sparsity of the natural image.

To solve this function, we use ISTA to convert it into an iterative manner. Then, the solution is
\begin{equation}
	\label{eq2}
	I^{SR}_{k+1}=\mathcal{T}_{\lambda\alpha_k}(I^{SR}_k-\alpha_k D^T(DI^{SR}_k-I^{LR})),
\end{equation}
where $\alpha_k$ is the weighting factor for the $k$-th iteration and $\mathcal{T}(\cdot)$ is the soft-thresholding operation.

It can be found that the right hand side of Equation \ref{eq2} has two independent variables $I^{SR}_k$ and $I^{LR}$. To make it clear for understanding, we re-write Equation \ref{eq2} as
\begin{equation}
	\label{eq3}
	I^{SR}_{k+1}=\mathcal{T}_{\lambda\alpha_k}((E-\alpha_k D^TD)I^{SR}_k-\alpha_k D^TI^{LR}),
\end{equation}
where $E$ is the identity matrix.

In Equation \ref{eq3}, we can find that $D^TI^{LR}$ is shared for every iteration. In this point of view, we can calculate this term before the ISTA optimization, and regard it as an invariant to speed up the optimization.

\subsection{Network Design}

Figure~\ref{fig:network} shows the entire network design of our ISTAR. Firstly, the input image $I^{LR}$ is converted into the feature space by one convolutional layer as
\begin{equation}
    \label{eq4}
    \hat{I}^{LR} = Conv(I^{LR}).
\end{equation}
Then, two convolutional layers and one ReLU activation are utilized to calculate the $D^TI^{LR}$ for ISTA steps, as shown in the figure. There are $K$ steps for ISTA optimization. For the $k$-th step, there is
\begin{equation}
    \label{eq5}
    \hat{I}^{SR}_k=ISTABlock(D^TI^{LR}, \hat{I}^{SR}_{k-1}),
\end{equation}
where $ISTABlock(\cdot)$ is the designed ISTA block.

After $K$ iterations, the output $\hat{I}^{SR}_K$ and $\hat{I}^{LR}$ are organized in a skip connection manner as
\begin{equation}
    \label{eq6}
    \hat{I}^{SR} = \hat{I}^{LR} + Padding(\hat{I}^{SR}_K),
\end{equation}
where $Padding(\cdot)$ aims to introduce the non-linearity. The padding structure is composed of two convolutional layers and one ReLU activation. Finally, the restored image $I^{SR}$ is generated from the SR feature $\hat{I}^{SR}$ as
\begin{equation}
    \label{eq7}
    I^{SR} = Upscaling(\hat{I}^{SR}),
\end{equation}
where $Upscaling(\cdot)$ is the upscaling module. The upscaling module is composed of one convolutional layer and a sub-pixel convolution.

\subsection{Design of ISTA Block}
\begin{figure}[t]
    \centering
    \includegraphics[width=0.75\linewidth]{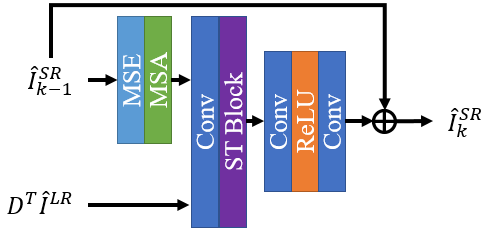}
    \caption{ISTA Block design}
    \label{fig:istablock}
\end{figure}

Figure~\ref{fig:istablock} shows the ISTA block design. The inputs of $k$-th ISTA block are $\hat{I}^{SR}_{k-1}$ and $D^TI^{LR}$, and the output of the block is $\hat{I}^{SR}_k$. Multi-scale exploration (MSE) block and multi-scale attention (MSA) mechanism are utilized to generate $(E-\alpha_k D^TD)\hat{I}^{SR}_k$ from $\hat{I}$. Then, one $1\times1$ convolution combines the information from the $(E-\alpha_k D^TD)\hat{I}^{SR}_k$ and the $D^TI^{LR}$, and the soft-thresholding block (ST Block) jointly explores the features following the Equation \ref{eq3}. A padding structure is introduced to the ISTA block with skip connection for better gradient transmission. The padding structure is composed of two convolutional layers and a ReLU activation.

\begin{figure}[t]
    \centering
    \includegraphics[width=0.75\linewidth]{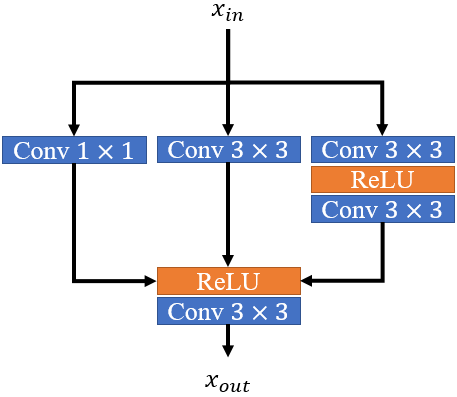}
    \caption{Multi-scale exploration (MSE) block design}
    \label{fig:mse}
\end{figure}

Figure \ref{fig:mse} shows the block design of MSE. Multi-scale design has proved to be an effective structure for image SR. In this block, three different scales ($1\times1$, $3\times3$, and $5\times5$) are considered to explore the hierarchical information from the features. To make the exploration more efficient, the $5\times5$ exploration is separated into two $3\times3$ convolutions with a ReLU activation, which hold the same receptive field. After the multi-scale feature extraction, one ReLU activation and a convolutional layer concatenate the hierarchical features and generate the final output of the MSE.

\begin{figure}[t]
    \centering
    \includegraphics[width=0.75\linewidth]{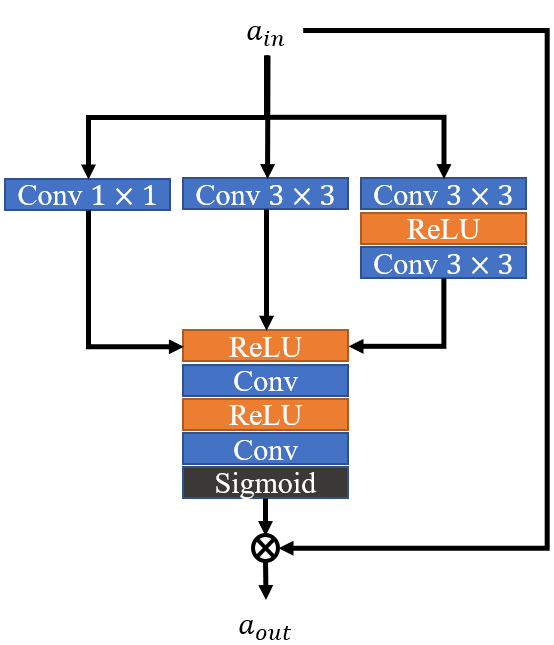}
    \caption{Multi-scale attention (MSA) mechanism design}
    \label{fig:msa}
\end{figure}

Figure \ref{fig:msa} shows the block design of multi-scale attention (MSA) mechanism. It can be found that the MSA has a similar multi-scale exploration design as MSE. Hierarchical features from three different scales are jointly explored by two convolutional layers and two ReLU activation operations. Then, the Sigmoid activation is introduced to generate the non-negative attention information. 

\begin{figure}[t]
    \centering
    \includegraphics[width=0.75\linewidth]{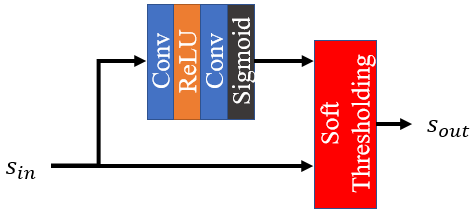}
    \caption{Soft thresholding block (ST Block) design}
    \label{fig:stblock}
\end{figure}

Figure \ref{fig:stblock} shows the soft-thresholding block (ST Block) design. The ST Block is composed of two $1\times1$ convolutional layers, one ReLU activatiion and a Sigmoid activation. The ST Block calculates the hyper parameters from the input features by the network design, and performs the soft-thresholding action with the learned parameters.

\subsection{Network Details}
In the network, all convolution kernels are set with size $3\times3$ except for the MSE and the ST Block. The convolutions in ST Block are set with kernel size as $1\times1$. The filters of all convolutional layers are set with 64 except for the ST Block and the Upscaling module. There are $K=16$ ISTA blocks in the ISTAR. The loss function is chosen as $\ell_1$-norm between the SR and HR images.

\section{Experiment}
\subsection{Settings}
The network is trained with DIV2K~\cite{div2k} dataset, which contains 900 high resolution images. We choose first 800 images for training, and last 5 image for validation. Five common benchmarks are chosen for comparing the restoration effectiveness: Set5~\cite{set5}, Set14~\cite{set14}, B100~\cite{b100}, Urban100~\cite{urban100} and Manga109~\cite{manga109}. We update our ISTAR for 1000 epochs by the Adam~\cite{adam} optimizer with learning rate $10^{-4}$. The learning rate is halved for every 200 epochs. The scaling factors are chosen as $\times2$, $\times3$ and $\times4$. The patch size for training is chosen as $48\times48$ for LR images. All other settings are same as RDN~\cite{rdn}. The objective indicators are chosen as peak signal-to-noise ratio~\cite{psnr} (PSNR) and structural similarity~\cite{ssim} (SSIM).

\subsection{Model Analysis}
\subsubsection{Investigation on the Iteration Times}

\begin{table}[t]
    \centering
    \caption{PSNR/SSIM comparisons among different iteration times with scaling factor $\times4$.}
    \label{tab:abl_K}
    \begin{tabular}{|c|c|c|c|c|c}
        \hline
         $K$& Set14& B100& Urban100& Manga109\\
         \hline
         4& 	28.46/0.7786&	27.47/0.7330&	25.75/0.7748&	30.06/0.9020\\
         8& 	28.50/0.7810&	27.53/0.7352&	25.95/0.7827&	30.33/0.9063\\
         12& 	28.57/0.7809&	27.57/0.7352&	26.04/0.7842&	30.46/0.9074\\
         16& 	28.59/0.7822&	27.59/0.7372&	26.11/0.7886&	30.54/0.9089\\
         \hline
    \end{tabular}
\end{table}

According to the ISTA optimization, the result becomes more accurate with the increase of iteration times. To investigate the effectiveness of the iteration times, we compare the performances with different iteration time $K$. Table \ref{tab:abl_K} shows the PSNR/SSIM comparisons among different $K$ with scaling factor $\times4$. For a fair comparison, all the testing models are updated for 200 epochs under the same settings. In the table, we can find that the PSNR and SSIM rises with the increase of iteration time $K$. When $K=16$, the network achieves the best performance. In this point of view, a deeper network leads to a better performance. When $K$ increases from 12 to 16, the PSNR and SSIM gets small improvement. To balance the performance and the computation complexity, we choose $K=16$ to build the ISTAR.

\subsubsection{Investigation on the Multi-Scale Exploration}
\begin{table}[t]
    \centering
    \caption{PSNR/SSIM comparisons among multi-scale explorations with scaling factor $\times4$.}
    \label{tab:abl_S}
    \begin{tabular}{|c|c|c|c|c}
        \hline
         $S$ &  B100& Urban100& Manga109\\
         \hline
         $1$        & 	27.55/0.7354&	26.00/0.7835&	30.40/0.9067\\
         $(1+3)$    & 	27.55/0.7355&	26.03/0.7844&	30.49/0.9073\\
         $(1+3+5)$  & 	27.59/0.7372&	26.11/0.7886&	30.54/0.9089\\
         \hline
    \end{tabular}
\end{table}
In MSE, we utilize three different scales to explore the hierarchical information. To investigate the effectiveness of multi-scale exploration, we conduct the experiments with scaling combination as $1$, $(1+3)$ and $(1+3+5)$. Table \ref{tab:abl_S} shows the PSNR/SSIM comparisons among different scales combinations. In the figure, we can find that the multi-scale exploration brings 0.08 dB PSNR improvement on Urban100 dataset and 0.05 db improvement on Manga109 dataset. Compared with $S=1$, the combination $(1+3)$ brings 0.09 dB PSNR improvement on Manga109 dataset. In this point of view, the multi-scale exploration is an effective design for restoration.

\subsubsection{Investigation on the Multi-Scale Attention}
\begin{table}[t]
    \centering
    \caption{PSNR/SSIM comparisons between MSA with scaling factor $\times4$.}
    \label{tab:abl_msa}
    \begin{tabular}{|c|c|c|c|c|c|}
        \hline
         MSA& Set14& B100& Urban100& Manga109\\
         \hline
         w/o& 	28.67/0.7874&	27.64/0.7388&	26.32/0.7942&	30.77/0.9120\\
         w&	    28.74/0.7855&	27.66/0.7391&	26.38/0.7955&	30.87/0.9130\\
         \hline
    \end{tabular}
\end{table}
In ISTA block, MSA is considered for better network representation. To show the effectiveness of MSA, we compare the objective performances with and without MSA on different benchmarks. Table \ref{tab:abl_msa} shows the PSNR/SSIM comparisons after training 1000 epochs. In the figure, we can find that the network with MSA has 0.1 dB improvement on Manga109 dataset and 0.06 dB improvement on Set14 and Urban100 dataset. In this point of view, the MSA proves to be an effective component for image restoration and improves the network capacity.

\subsection{Results}

\begin{table*}[!ht]
	\centering
	\caption{PSNR/SSIM comparisons with scaling factor $\times2$, $\times3$, and $\times4$ on five benchmarks. The best performances are shown in \textbf{bold}.}
	\label{tab:BI-result}
	\begin{tabular}{|c|c|c|c|c|c|c|}
		\hline
		\multirow{2}{*}{Scale}& \multirow{2}{*}{Model}&  
		Set5~\cite{set5}& Set14~\cite{set14}& B100~\cite{b100}& Urban100~\cite{urban100}& Manga109~\cite{manga109} \\
		& & PSNR/SSIM & PSNR/SSIM & PSNR/SSIM & PSNR/SSIM & PSNR/SSIM\\
		\hline
		\multirow{18}{*}{$\times2$} &SRCNN~\cite{srcnn}&
		36.66 / 0.9542& 32.42 / 0.9063& 31.36 / 0.8879& 29.50 / 0.8946& 35.74 / 0.9661\\
		
		& FSRCNN~\cite{fsrcnn}&
		37.00 / 0.9558& 32.63 / 0.9088& 31.53 / 0.8920& 29.88 / 0.9020& 36.67 / 0.9694\\
		
		& VDSR~\cite{vdsr}&
		37.53 / 0.9587& 33.03 / 0.9124& 31.90 / 0.8960& 30.76 / 0.9140& 37.22 / 0.9729\\
		
		& DRCN~\cite{drcn}&
		37.63 / 0.9588& 33.04 / 0.9118& 31.85 / 0.8942& 30.75 / 0.9133& 37.63 / 0.9723\\
		
		& CNF~\cite{cnf}&
		37.66 / 0.9590& 33.38 / 0.9136& 31.91 / 0.8962& - & - \\
		
		&LapSRN~\cite{lapsrn}&
		37.52 / 0.9590& 33.08 / 0.9130& 31.80 / 0.8950& 30.41 / 0.9100& 37.27 / 0.9740\\
		
		&DRRN~\cite{drrn}&
		37.74 / 0.9591& 33.23 / 0.9136& 32.05 / 0.8973& 31.23 / 0.9188& 37.92 / 0.9760\\
		
		&BTSRN~\cite{btsrn}&
		37.75 / -& 33.20 / -& 32.05 / -& 31.63 / -& -\\
		
		&MemNet~\cite{memnet}&
		37.78 / 0.9597& 33.28 / 0.9142& 32.08 / 0.8978& 31.31 / 0.9195& 37.72 / 0.9740 \\
		
		&SelNet~\cite{selnet}&
		37.89 / 0.9598& 33.61 / 0.9160& 32.08 / 0.8984& - &  - \\
		
		&CARN~\cite{carn}&
		37.76 / 0.9590& 33.52 / 0.9166& 32.09 / 0.8978& 31.92 / 0.9256& 38.36 / 0.9765\\
		
		&IMDN~\cite{imdn}&
		38.00 / 0.9605& 33.63 / 0.9177& 32.19 / 0.8996& 32.17 / 0.9283& 38.88 / 0.9774\\
		
		&RAN~\cite{ran}&
		37.58 / 0.9592 &33.10 / 0.9133 &31.92 / 0.8963& -& -\\
		
		&DNCL~\cite{dncl}&
		37.65 / 0.9599 &33.18 / 0.9141 &31.97 / 0.8971 &30.89 / 0.9158& - \\
		
		&FilterNet~\cite{filternet}&
		37.86 / 0.9610 &33.34 / 0.9150 &32.09 / 0.8990 &31.24 / 0.9200& - \\
		
		&MRFN~\cite{mrfn}&
		37.98 / 0.9611 &33.41 / 0.9159 &32.14 / 0.8997 &31.45 / 0.9221 &38.29 / 0.9759\\
		
		&SeaNet-baseline~\cite{seanet}&
		37.99 / 0.9607 & 33.60 / 0.9174 & 32.18 / 0.8995 & 32.08 / 0.9276 & 38.48 / 0.9768 \\
		
		&DEGREE~\cite{degree}&
		37.58 / 0.9587& 33.06 / 0.9123& 31.80 / 0.8974& - & - \\

		&Cross-SRN~\cite{liu2021_tcsvt}&
		38.03 / 0.9606 & 33.62 / 0.9180 & 32.19 / 0.8997 & 32.28 / 0.9290 & 38.75 / 0.9773 \\

		&ISTAR (Ours)& 
		\textbf{38.15 / 0.9610}&	\textbf{33.79 / 0.9197}&	\textbf{32.29 / 0.9010}&	\textbf{32.65 / 0.9331}&	\textbf{38.96 / 0.9777}\\
		
		\hline
		
		\multirow{17}{*}{$\times3$}& SRCNN~\cite{srcnn} &
		32.75 / 0.9090& 29.28 / 0.8209& 28.41 / 0.7863& 26.24 / 0.7989& 30.59 / 0.9107\\
		
		&FSRCNN~\cite{fsrcnn}&
		33.16 / 0.9140& 29.43 / 0.8242& 28.53 / 0.7910& 26.43 / 0.8080& 30.98 / 0.9212\\
		
		&VDSR~\cite{vdsr}&
		33.66 / 0.9213& 29.77 / 0.8314& 28.82 / 0.7976& 27.14 / 0.8279& 32.01 / 0.9310\\
		
		&DRCN~\cite{drcn}&
		33.82 / 0.9226& 29.76 / 0.8311& 28.80 / 0.7963& 27.15 / 0.8276& 32.31 / 0.9328\\
		
		&CNF~\cite{cnf}&
		33.74 / 0.9226& 29.90 / 0.8322& 28.82 / 0.7980& - & - \\
		
		&DRRN~\cite{drrn}&
		34.03 / 0.9244& 29.96 / 0.8349& 28.95 / 0.8004& 27.53 / 0.8378& 32.74 / 0.9390\\
		
		&BTSRN~\cite{btsrn}&
		34.03 / -& 29.90 / -& 28.97 / -& 27.75 / -& -\\
		
		&MemNet~\cite{memnet}&
		34.09 / 0.9248& 30.00 / 0.8350& 28.96 / 0.8001& 27.56 / 0.8376& 32.51 / 0.9369 \\
		
		&SelNet~\cite{selnet}&
		34.27 / 0.9257& 30.30 / 0.8399& 28.97 / 0.8025& - &  - \\
		
		&CARN~\cite{carn}&
		34.29 / 0.9255& 30.29 / 0.8407& 29.06 / 0.8034& 28.06 / 0.8493& 33.50 / 0.9440\\
		
		&IMDN~\cite{imdn}&
		34.36 / 0.9270& 30.32 / 0.8417& 29.09 / 0.8046& 28.17 / 0.8519& 33.61 / 0.9445\\
		
		&RAN~\cite{ran}&
		33.71 / 0.9223 & 29.84 / 0.8326 & 28.84 / 0.7981 & - & -\\
		
		&DNCL~\cite{dncl}&
		33.95 / 0.9232 & 29.93 / 0.8340 & 28.91 / 0.7995 & 27.27 / 0.8326 & -\\
		
		&FilterNet~\cite{filternet}&
		34.08 / 0.9250 & 30.03 / 0.8370 & 28.95 / 0.8030 & 27.55 / 0.8380 & -\\
		
		&MRFN~\cite{mrfn}&
		34.21 / 0.9267 & 30.03 / 0.8363 & 28.99 / 0.8029 & 27.53 / 0.8389 & 32.82 / 0.9396 \\
		
		&SeaNet-baseline~\cite{seanet}&
		34.36 / 0.9280 & 30.34 / 0.8428 & 29.09 / 0.8053 & 28.17 / 0.8527 & 33.40 / 0.9444 \\
		
		&DEGREE~\cite{degree}&
		33.76 / 0.9211& 29.82 / 0.8326& 28.74 / 0.7950& - & - \\
		
		&Cross-SRN&
		34.43 / 0.9275& 30.33 / 0.841& 29.09 / 0.8050& 28.23 / 0.8535& 33.65 / 0.9448\\
		
		&ISTAR~(Ours)&
		\textbf{34.55 / 0.9284}&	\textbf{30.48 / 0.8452}&	\textbf{29.18 / 0.8076}&	\textbf{28.56 / 0.8610}&	\textbf{33.85 / 0.9467}\\
		
		\hline
		\multirow{19}{*}{$\times4$}&SRCNN~\cite{srcnn}&
		30.48 / 0.8628& 27.49 / 0.7503& 26.90 / 0.7101& 24.52 / 0.7221& 27.66 / 0.8505\\
		
		&FSRCNN~\cite{fsrcnn}&
		30.71 / 0.8657& 27.59 / 0.7535& 26.98 / 0.7150& 24.62 / 0.7280& 27.90 / 0.8517\\
		
		&VDSR~\cite{vdsr}&
		31.35 / 0.8838& 28.01 / 0.7674& 27.29 / 0.7251& 25.18 / 0.7524& 28.83 / 0.8809\\
		
		&DRCN~\cite{drcn}&
		31.53 / 0.8854& 28.02 / 0.7670& 27.23 / 0.7233& 25.14 / 0.7510& 28.98 / 0.8816\\
		
		&CNF~\cite{cnf}&
		31.55 / 0.8856& 28.15 / 0.7680& 27.32 / 0.7253& - & - \\
		
		&LapSRN~\cite{lapsrn}&
		31.54 / 0.8850& 28.19 / 0.7720& 27.32 / 0.7280& 25.21 / 0.7560& 29.09 / 0.8845\\
		
		&DRRN~\cite{drrn}&
		31.68 / 0.8888& 28.21 / 0.7720& 27.38 / 0.7284& 25.44 / 0.7638& 29.46 / 0.8960\\
		
		&BTSRN~\cite{btsrn}&
		31.85 / -& 28.20 / -& 27.47 / -& 25.74 / -& -\\
		
		&MemNet~\cite{memnet}&
		31.74 / 0.8893& 28.26 / 0.7723& 27.40 / 0.7281& 25.50 / 0.7630& 29.42 / 0.8942 \\
		
		&SelNet~\cite{selnet}&
		32.00 / 0.8931& 28.49 / 0.7783& 27.44 / 0.7325& - &  - \\
		
		&SRDenseNet~\cite{srdensenet}&
		32.02 / 0.8934& 28.50 / 0.7782& 27.53 / 0.7337& 26.05 / 0.7819& - \\
		
		&CARN~\cite{carn}&
		32.13 / 0.8937& 28.60 / 0.7806& 27.58 / 0.7349& 26.07 / 0.7837& 30.47 / 0.9084\\
		
		&IMDN~\cite{imdn}&
		32.21 / 0.8948& 28.58 / 0.7811& 27.56 / 0.7353& 26.04 / 0.7838& 30.45 / 0.9075\\
		
		&RAN~\cite{ran}&
		31.43 / 0.8847 & 28.09 / 0.7691 & 27.31 / 0.7260 & - & -\\
		
		&DNCL~\cite{dncl}&
		31.66 / 0.8871 & 28.23 / 0.7717 & 27.39 / 0.7282 & 25.36 / 0.7606 & -\\
		
		&FilterNet~\cite{filternet}&
		31.74 / 0.8900 & 28.27 / 0.7730 & 27.39 / 0.7290 & 25.53 / 0.7680 & -\\
		
		&MRFN~\cite{mrfn}&
		31.90 / 0.8916 & 28.31 / 0.7746 & 27.43 / 0.7309 & 25.46 / 0.7654 & 29.57 / 0.8962 \\
		
		&SeaNet-baseline~\cite{seanet}&
		32.18 / 0.8948 & 28.61 / 0.7822 & 27.57 / 0.7359 & 26.05 / 0.7896 & 30.44 / 0.9088 \\
		
		&DEGREE~\cite{degree}&
		31.47 / 0.8837& 28.10 / 0.7669& 27.20 / 0.7216& - & - \\

		&Cross-SRN~\cite{liu2021_tcsvt}&
		{32.24 / 0.8954}& {28.59} / {0.7817}& {27.58 / 0.7364}& {26.16 / 0.7881}& {30.53} / {0.9081}\\
		
		&ISTAR (Ours)&
		\textbf{32.37 / 0.8972}&	\textbf{28.74 / 0.7855}&	\textbf{27.66 / 0.7391}&	\textbf{26.38 / 0.7955}&	\textbf{30.87 / 0.9130}\\
		
		\hline
	\end{tabular}
\end{table*}

We compare our ISTAR with several traditional and recent CNN-based image SR works: SRCNN~\cite{srcnn}, FSRCNN~\cite{fsrcnn}, VDSR~\cite{vdsr}, DRCN~\cite{drcn}, CNF~\cite{cnf}, LapSRN~\cite{lapsrn}, DRRN~\cite{drrn}, BTSRN~\cite{btsrn}, MemNet~\cite{memnet}, SelNet~\cite{selnet}, SRDenseNet~\cite{srdensenet}, CARN~\cite{carn}, IMDN~\cite{imdn}, RAN~\cite{ran}, DNCL~\cite{dncl}, FilterNet~\cite{filternet}, MRFN~\cite{mrfn}, SeaNet~\cite{seanet}, DEGREE~\cite{degree} and Cross-SRN~\cite{crosssrn}. Table~\ref{tab:BI-result} shows the PSNR/SSIM comparisons on five testing benchmarks with scaling factor $\times2$, $\times3$ and $\times4$.

In the figure, we can find that our ISTAR achieves the best performance on all testing benchmarks with all scaling factors. Compared with Cross-SRN, our ISTAR achieves near 0.4 dB, 0.3 dB and 0.2 dB improvement on Urban100 dataset with scaling factor $\times2$, $\times3$ and $\times4$ separately. When sacling factor is $\times4$, ISTAR achieves 0.34 dB PSNR higher than Cross-SRN on Manga109 dataset. It should be noticed that Urban100 and Manga109 are two representative datasets with plentiful edges and lines. In this point of view, ISTAR can effectively restore the structural information than other works.

\begin{table}
		\centering
		\caption{PSNR/SSIM, parameters and MACs comparisons with optimization-inspired networks with scaling factor $\times2$.}
		\label{tab:hqs}
		\fontsize{6.5}{8}\selectfont
		\begin{tabular}{|c|c|c|c|c|}
			\hline
			\textbf{Method}& \textbf{DBPN~\cite{dbpn}}& \textbf{USRNet~\cite{usrnet}}	& \textbf{HSRNet~\cite{hsrnet}}& ISTAR (Ours) \\
			\hline
			\hline
			\textbf{Param(M)}&  5.95& 17.01&  1.26&   5.05\\
			\textbf{MACs(G)}&   3746.2& 8545.8&  808.2& 1164.31 \\
			\hline
			\hline
			\textbf{Set5}&	    38.09 / 0.9600& 37.76/0.9599& 38.07 / 0.9607& 38.15 / 0.9610\\ 
			\textbf{Set14}&		33.85 / 0.9190& 33.43/0.9159& 33.78 / 0.9197& 33.79 / 0.9197\\
			\textbf{B100}& 		32.27 / 0.9000& 32.09/0.8985& 32.26 / 0.9006& 32.29 / 0.9010\\
			\textbf{Urban100}& 	32.55 / 0.9324& 31.78/0.9259& 32.53 / 0.9320& 32.65 / 0.9331\\
			\hline
		\end{tabular}
	\end{table}

To further investigate the effectiveness of ISTAR, we compare the network with several optimization-inspired methods. Table \ref{tab:hqs} shows the PSNR/SSIM comparisons with differnet optimization-inspired networks. DBPN~\cite{dbpn} is developed by the iterative back projection algorithm, while USRNet~\cite{usrnet} and HSRNet~\cite{hsrnet} are inspired by the half-quadratic splitting (HQS) strategy. The MACs is calculated by the same method as the HSRNet. In the table, we can find that our ISTAR acheives better performance than USRNet and HSRNet. Compared with HSRNet, our method achieves 0.1 dB PSNR improvement on Set5 and Urban100 datasets. Furthrmore, ISTAR achieves better performance than USRNet with near 29.7\% parameters and 13.7\% computation complexity. DBPN is one of the state-of-the-art image SR methods. Compared with DBPN, our method achieves competitive or better PSNR/SSIM results with similar parameters and 68.9\% MACs off. In this point of view, ISTAR proves to be an effective optimization scheme for image SR.

\begin{figure*}[t]
    \centering
    \includegraphics[width=.6\linewidth]{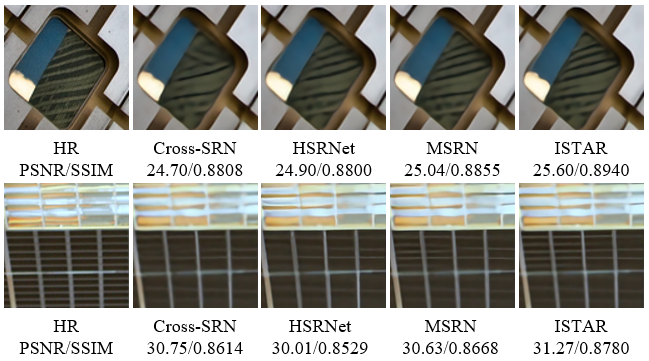}
    \caption{Visualization comparisons on Urban100 dataset with scaling factor $\times4$}
    \label{fig:urban100}
\end{figure*}

Figure \ref{fig:urban100} shows the visualization comparisons with different recnet image SR works (Cross-SRN~\cite{crosssrn}, HSRNet~\cite{hsrnet} and MSRN~\cite{msrn}) on Urban100 dataset with scaling factor $\times4$. In the first row of the figure, we can find that the building restored by ISTAR is closest to the ground-truth than other methods. Similarly, in the second row of the figure, ISTAR can restore the lines and girds more effectively than other works with much higher PSNR/SSIM results. In this point of view, ISTAR can generate more satisfying subjective results than other works.

\section{Conclusion}
In this paper, we proposed an ISTA-inspired restoration network, termed ISTAR, for effective image super-resolution. We analyzed the image super-resolution task from the optimization perspective and proposed an iterative solution based on the ISTA. According to the formulation of the solution, an end-to-end network with optimization-inspired blocks was developed for effective image super-resolution. Multi-scale exploration and multi-scale attention mechanism were specifically devised to boost the network capacity. Experimental results show the proposed ISTAR achieves better subjective and objective performances than other state-of-the-art works. Compared with other optimization-inspired methods, ISTAR achieves competitive or better performance with much fewer parameters and lower computation complexity.

\bibliographystyle{IEEEtran}
\bibliography{main}

\end{document}